\documentclass[twocolumn]{aastex631}
\usepackage{comment}
\usepackage{amsmath}
\usepackage{soul}

\shorttitle{Magnetic fields in solar plage regions} %AASTeX v6.3.1 Sample article}
\shortauthors{da Silva Santos et al.}
\DeclareRobustCommand{\ion}[2]{\textup{#1\,\textsc{\lowercase{#2}}}}

\graphicspath{{./}{figures/}}
\newcommand{\uat}[2]{\href{http://astrothesaurus.org/uat/#2}{#1 (#2)}}

\begin{document}

\title{Magnetic fields in solar plage regions: insights from high-sensitivity spectropolarimetry}

\correspondingauthor{Jo\~{a}o Santos}
\email{jdasilvasantos@nso.edu}
% https://journals.aas.org/aastexguide/

\author[0000-0002-3009-295X]{J. M. da Silva Santos}
\affiliation{National Solar Observatory, 3665 Discovery Drive, Boulder, CO 80303, USA}
%\collaboration{6}{(AAS Journals Data Editors)}

\author[0000-0001-8016-0001]{K. Reardon}
\affiliation{National Solar Observatory, 3665 Discovery Drive, Boulder, CO 80303, USA}
\affiliation{Department of Astrophysics and Planetary Sciences, University of Colorado, Boulder, CO 80303, USA}

\author[0000-0002-6116-7301]{G. Cauzzi}
\affiliation{National Solar Observatory, 3665 Discovery Drive, Boulder, CO 80303, USA}

\author[0000-0002-7451-9804]{T. Schad}
\affiliation{National Solar Observatory, 22 \'{}\={O}hi\'{}a K\={u} Street,
Pukalani, HI 96768, USA}

\author[0000-0001-7764-6895]{V. Martínez Pillet}
\affiliation{National Solar Observatory, 3665 Discovery Drive, Boulder, CO 80303, USA}

\author[0000-0003-3147-8026]{A. Tritschler} 
\affiliation{National Solar Observatory, 3665 Discovery Drive, Boulder, CO 80303, USA}

\author{F. Wöger}
\affiliation{National Solar Observatory, 3665 Discovery Drive, Boulder, CO 80303, USA}

\author[0000-0002-5556-6840]{R. Hofmann}
\affiliation{National Solar Observatory, 3665 Discovery Drive, Boulder, CO 80303, USA}
\affiliation{Department of Astrophysics and Planetary Sciences, University of Colorado, Boulder, CO 80303, USA}

\author[0000-0002-6495-4685]{J. Stauffer}
\affiliation{National Solar Observatory, 3665 Discovery Drive, Boulder, CO 80303, USA}
\affiliation{Department of Astrophysics and Planetary Sciences, University of Colorado, Boulder, CO 80303, USA}

\author[0000-0002-2554-1351]{H. Uitenbroek}
\affiliation{National Solar Observatory, 3665 Discovery Drive, Boulder, CO 80303, USA}

\begin{abstract}
 
Plage regions are patches of concentrated magnetic field in the Sun's atmosphere where hot coronal loops are rooted.  While previous studies have shed light on the properties of plage magnetic fields in the photosphere, there are still challenges in measuring the overlying chromospheric magnetic fields, which are crucial to understanding the overall heating and dynamics. Here, we utilize high-sensitivity, spectropolarimetric data obtained by the four-meter Daniel K. Inouye Solar Telescope (DKIST) to investigate the dynamic environment and magnetic field stratification of an extended, decaying plage region.
The data show strong circular polarization signals in both plage cores and surrounding fibrils. Notably, weak linear polarization signals clearly differentiate between plage patches and the fibril canopy, where they are relatively stronger. Inversions of the \ion{Ca}{II} 8542\,\AA~spectra show an imprint of the fibrils in the chromospheric magnetic field, with typical field strength values ranging from $\sim$\,$200-300$\,G in fibrils.
We confirm the weak correlation between field strength and cooling rates in the lower chromosphere. Additionally, we observe supersonic downflows and strong velocity gradients in the plage periphery, indicating dynamical processes occurring in the chromosphere. These findings contribute to our understanding of the magnetic field and dynamics within plages, emphasizing the need for further research to explore the expansion of magnetic fields with height and the three-dimensional distribution of heating rates in the lower chromosphere. 
\end{abstract}



\section{Introduction} 
\label{sec:intro}

Solar plages are extended magnetized regions, initially characterized by their enhanced emission in the integrated \ion{Ca}{II} H and K resonance lines \citep[e.g.,][]{1974SoPh...39...49S, 1975ApJ...200..747S}. 
They are commonly observed in the proximity of pores and sunspots within active regions (ARs), serving as a source region of hot coronal loops. Plages can persist as ARs remnants long after the cessation of flux emergence and the decay of sunspots, significantly contributing to the ultraviolet solar irradiance \citep[e.g.,][]{1998ApJ...492..390L}. %Eventually, plages dissipate through the cancellation of magnetic flux with opposite polarity and outward diffusion \citep[e.g.,][]{1986ApJ...306..304T,2000ssma.book.....S}. 
Central to understanding the chromosphere in plages is the study of their magnetic fields, which are recognized as key drivers of dynamics and heating in this region \citep[see review by ][]{2019ARA&A..57..189C}.

It is well established that plage regions harbor concentrations of predominantly unipolar, fairly vertical (10$^\circ - 15^\circ$) kilo-gauss magnetic fields in the photosphere \citep[e.g.,][]{1992ApJ...390L.103R,1992ApJ...396..351T,1997ApJ...474..810M,2010A&A...524A...3N,2015A&A...576A..27B}. 
The magnetic concentrations expand in the upper photosphere, displaying more inclined and weaker fields \citep[e.g.,][]{1989ApJ...340..571Z,1995A&A...293..240B,2015A&A...576A..27B}. 

Measuring plage magnetic fields in the chromosphere, however, has been a significant challenge. 
The formation mechanisms of chromospheric spectral lines are quite complex, making their interpretation difficult;  additionally, their limited magnetic sensitivity (small Landé factors, large Doppler broadening) along with the relatively weak field strengths in the chromosphere results in polarization signals that are at most a few $10^{-3}$ relative to the continuum \citep[e.g., in the \ion{Ca}{II} lines, ][]{2017SSRv..210..109D}, posing a difficult observational problem.

Nevertheless, a few recent papers have provided insights into chromospheric magnetic fields in plages. 
Using SST/CRISP spectropolarimetry, 
\citet{2020A&A...642A.210M} used the weak field approximation (WFA) in the \ion{Ca}{II} 8542\,\AA~line to derive the line-of-sight (LOS) component of the magnetic field ($B_{\rm LOS}$) in a small plage at disk center. They found that the average $B_{\rm LOS}$ in the plage was $\sim$\,400$-$450\,G at chromospheric heights ($\ge$\,1100 km). Despite the line core showing numerous chromospheric fibrils originating from the plage, the spatial distribution of the magnetic field appeared smooth and volume-filling.
Also using SST/CRISP, \citet{2020A&A...644A..43P} employed nonlocal thermodynamic equilibrium (NLTE) inversions and obtained similar results regarding the field properties, with the field inclination in plage similar to typical photospheric values ($\sim$\,10$^{\circ}$). While the overall magnetic field maps were fairly noisy, these authors could also measure the field in some of the chromospheric fibrils, finding higher inclinations ($\sim$\,50$^{\circ}$) and lower field strengths ($\sim$\,300\,G) than above the plage. %high-resolution (0.1$^{\prime\prime}$)

Chromospheric field strengths around 400$-$600\,G were inferred in AR plage by \citet{2021ApJ...921...39A} based on \ion{He}{I} 10830\,\AA~GREGOR/GRIS data. Interestingly, the inclination of the field was essentially horizontal to the surface in the small ($\sim$\,$6^{\prime\prime}\times 6^{\prime\prime}$) field of view (FOV) sampled. Finally, WFA inversions of CLASP2 observations of an AR plage yielded $B_{\rm LOS}$ values in the range $\sim$\,100$-$600\,G in the lower chromosphere (as derived from the \ion{Mn}{I} lines) but smaller values ($B_{\rm LOS}$\,$\sim$\,300\,G) at greater heights probed by \ion{Mg}{II} h and k \citep{2021SciA....7.8406I}. 

In this study, we further characterize the dynamic environment and magnetic field stratification of plage regions utilizing spectropolarimetric data obtained by the Daniel K. Inouye Solar Telescope \citep[DKIST,][]{2020SoPh..295..172R} during the operations commissioning phase. 

% ========================================================% 
\section{Data}
\label{sec:Observations}

\begin{figure}
    \centering
    \includegraphics[width=0.89\linewidth]{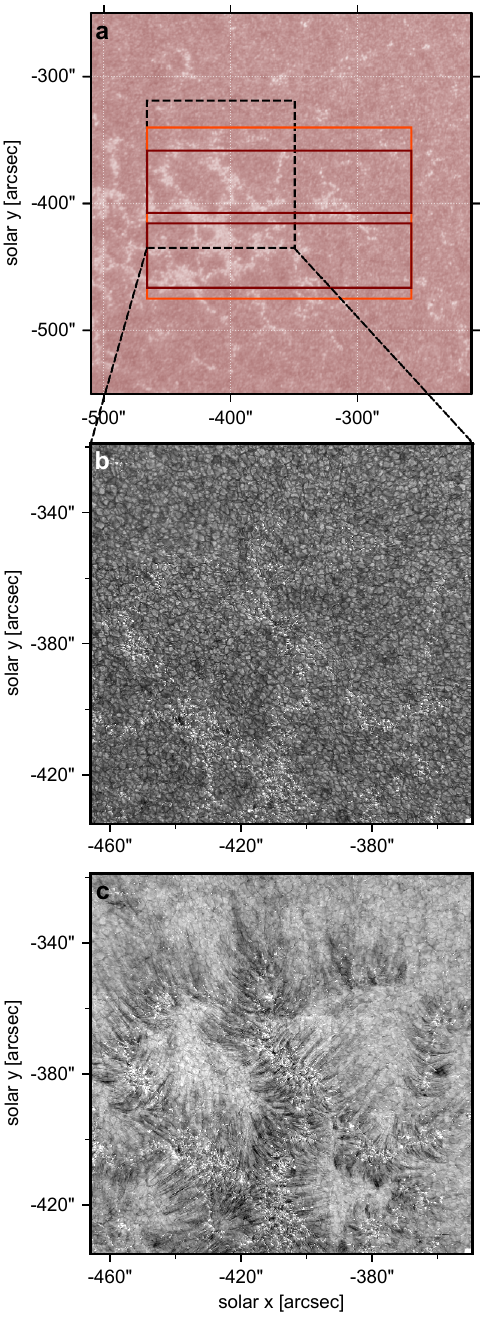}
    \caption{Overview of the target on 3 June 2022 at 17:46 UT. Panel a: SDO/AIA 1700\,\AA~intensities; the dashed black square shows one of the four VBI pointings composed of nine sub-fields each; the orange(red) rectangle shows the ViSP FOV for arm 6302\,\AA(8542\,\AA). Panel b: VBI G-band intensities. Panel c: VBI H$\beta$ intensities.}
    \label{fig:context}
\end{figure}

We present data taken with the Visible Spectro-Polarimeter \citep[ViSP,][]{2022SoPh..297...22D} and the Visible Broadband Imager \citep[VBI,][]{2021SoPh..296..145W} installed at DKIST. We also show line-of-sight magnetograms provided by the Helioseismic and Magnetic Imager \citep[HMI, ][]{2012SoPh..275..207S} and a UV continuum image acquired by the Atmospheric Imaging Assembly \citep[AIA,][]{2012SoPh..275...17L} onboard the Solar Dynamics Observatory \citep[SDO,][]{2012SoPh..275....3P}. The target was an extended, decaying plage region observed on 3 June 2022 between 17:09--20:59 UT (Fig. \ref{fig:context}\textcolor{xlinkcolor}{a}), identified as the likely source region of the solar wind properties measured in situ by the Parker Solar Probe \citep[PSP,][]{2016SSRv..204....7F} during Encounter \#12. The approximate central helioprojective coordinates were (-360$^{\prime\prime}$, -410$^{\prime\prime}$), spanning a range of $\mu$ (the cosine of the heliocentric angle) values between [0.7, 0.9]. 

In this paper, we used the Level 1 data that were publicly released\footnote{\url{https://nso.edu/dkist/data-center/}} in April 2023. The VBI data, used here for context, comprise speckle-reconstructed, high-resolution (0$^{\prime\prime}$.0106) images in the G-band, \ion{Ca}{II} K, and H$\beta$ filters, composing mosaics of four different copointings with each ViSP raster scans. The four VBI maps are each assembled from nine sub-fields to cover the whole FOV of $115^{\prime\prime}\times 117^{\prime\prime}$ (examples in Fig.\,\ref{fig:context}\textcolor{xlinkcolor}{b-c}). The magnetic elements constituting the plage are well visible as brightenings in AIA 1700\,\AA~(Fig.\,\ref{fig:context}\textcolor{xlinkcolor}{a}) and G-band imaging (Fig. \ref{fig:context}\textcolor{xlinkcolor}{b}), and identify the footpoints of the chromospheric fibril structures seen in H$\beta$ (Fig. \ref{fig:context}\textcolor{xlinkcolor}{c}). 

ViSP used two of the three spectrographs arms observing the \ion{Fe}{I} 6301/6302\,\AA~and \ion{Ca}{II} 8542\,\AA~lines. The data consist of eight, 490-step raster scans taking  $\sim$27\,min each. The eight scans comprised two repetitions of a four-tile mosaic covering at total area of 210$^{\prime\prime}$\,$\times$\,136$^{\prime\prime}$ for the 6302\,\AA~arm and 210$^{\prime\prime}$\,$\times$\,100$^{\prime\prime}$ for the 8542\,\AA~arm, with an overall mosaic cadence $\sim$113\,min. 
Based on comparisons with HMI continuum images, we determined the pixel scale along the slit to be 0$^{\prime\prime}$.0298 for the 6302\,\AA~arm and 0$^{\prime\prime}$.0194 for the 8542\,\AA~arm, with a slit length of 76.1$^{\prime\prime}$ and 50.3$^{\prime\prime}$ respectively (Fig.\,\ref{fig:context}\textcolor{xlinkcolor}{a}).
The wide ViSP slit (0$^{\prime\prime}$.214) and large raster step (0$^{\prime\prime}$.219) were chosen in order to sample a large portion of the  plage region in a reasonable time interval. The spectral dispersion is $\sim$0.0128\,\AA~for arm 6302\,\AA~and $\sim$0.0188\,\AA~for arm 8542\,\AA.
At each slit position, a sequence of 10 modulation states was repeated 12 times, for a total integration of 480 msec per pixel per slit position, and requiring 3.288 sec per slit position.

The ViSP data require additional post-processing prior to analysis. That includes a correction for residual crosstalk, wavelength/flux calibration, and arm coalignment. We detail the post-processing steps applied to the Level 1 data in the Appendix \ref{section:post_processing}. 

%\vfill\null % to force columnbreak

% ========================================================% 
\section{Methods}
\label{sec:Methods}

\begin{figure*}[ht]
    \centering
    \includegraphics[width=0.75\linewidth]{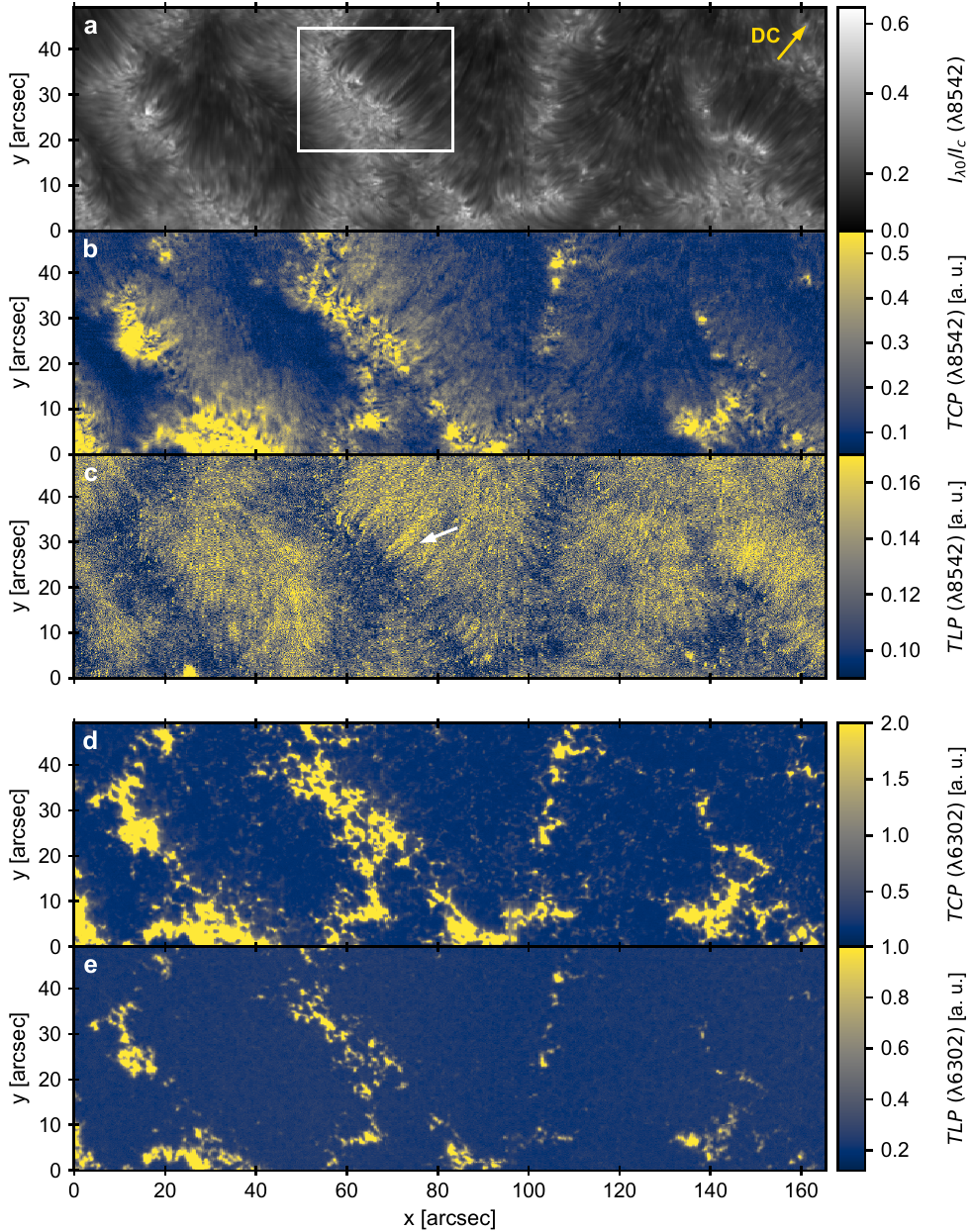}
    \caption{Polarization signals arising from the photosphere and chromosphere. Panel a: continuum-normalized intensity in the core of $\lambda$8542; the yellow arrow points to the disk center. Panels b/c: $\lambda$8542 TCP/TLP in arbitrary units. Panels d/e: TCP/TLP for $\lambda$6302. All color bars are capped for display purposes.}    
    \label{fig:TLPTCP1}
\end{figure*}

We performed Milne-Eddington (ME) inversions of the \ion{Fe}{I} lines (hereafter $\lambda$6302) using \texttt{PyMilne} \citep{2019A&A...631A.153D} -- a parallel Python code\footnote{\url{https://github.com/jaimedelacruz/pyMilne}} based on analytical response functions \citep{2007A&A...462.1137O}. 
We used the WFA code\footnote{\url{https://github.com/morosinroberta/spatial_WFA}} developed by \citet{2020A&A...642A.210M} for the \ion{Ca}{II} 8542\,\AA~line (hereafter $\lambda$8542). 
This algorithm is not only fast but it is also especially effective in mitigating noise through spatial regularization. The ViSP spectra were rebinned to a square pixel scale of 0$^{\prime\prime}$.219 prior to the inversions, resulting in a noise level of $\sim$8$\times10^{-4}$ relative to the continuum. The WFA is adequate for this line under most magnetic field regimes in the chromosphere, especially for determining $B_{\rm LOS}$; however, reliable estimates of the transverse component ($B_{\rm TRV}$) require a noise level less than $\sim$$10^{-3}$  and a location far enough from the limb to limit scattering polarization effects \citep[e.g.,][]{2010ApJ...722.1416M,2018ApJ...866...89C}. 
To avoid significant photospheric contamination, we only used the wavelength points within approximately $\pm$120 m\AA~from the line center, similar to \citet{2020A&A...642A.210M}. %(a few seconds on a personal laptop for the whole FOV)

We obtained LOS velocity maps in the chromosphere by computing the $\lambda8542$ line core Doppler shift. The algorithm uses the derivative of the intensity profile to identify the zero-crossings indicating the position of the absorption (or emission) cores. If the line shape deviates strongly from the common absorption profile, showing multiple intensity peaks, we take the wavelength position of the central peak. 
For $\lambda$6302, we obtained the LOS velocities from the ME inversions. %Post-processing was required to obtain smooth polarization maps and dopplergrams (Appendix \ref{section:polresiduals} and \ref{section:vlos}). 

The $\sim$12-18\AA-wide ViSP bandwidth includes other photospheric lines suitable for multi-line, NLTE inversions \citep{2022A&A...660A..37R}; those inversions are not only more computationally demanding than ME and WFA inversions, especially for large FOVs as in this paper, but also require very accurate intensity and polarization calibration; efforts to produce such Level 2 data are currently underway.

%Hanle effects can be ignored for field strengths above 80 G \citep{2021ApJ...918...15C}.

% ========================================================% 
\section{Results}
\label{section:results}

\subsection{Intensity and polarization features}
 
Figure \ref{fig:TLPTCP1} presents a comparison of polarization maps in the photosphere and chromosphere in the upper half of the FOV (top red rectangle in Fig.\,\ref{fig:context}). %The pixel scale is about a factor of 7(11) higher along the slit direction (y-axis) for $\lambda$6302(8542). 
In the plage, we find strong circular polarization signals in the photosphere (Fig.\,\ref{fig:TLPTCP1}\textcolor{xlinkcolor}{d}) and chromosphere (Fig.\,\ref{fig:TLPTCP1}\textcolor{xlinkcolor}{b}). Notably, the $\lambda$8542 total circular polarization ($TCP = \sum_{\lambda}|V_{\lambda}|/I_{\lambda}$) within $\pm$300\,$\rm m\mathring{A}$ from line center shows an imprint of the fibril canopy, extending up to $\sim$20$^{\prime\prime}$ from the plage footpoints (Fig. \ref{fig:TLPTCP1}\textcolor{xlinkcolor}{a-b}, also cf. Fig.\,\ref{fig:context}\textcolor{xlinkcolor}{c} ). The circular polarization signals in the chromosphere are more prominent towards the direction of the disk center due to projection effects. Comparison of TCP maps in the photosphere (Fig. \ref{fig:TLPTCP1}\textcolor{xlinkcolor}{e}) and chromosphere (Fig.\,\ref{fig:TLPTCP1}\textcolor{xlinkcolor}{b}) clearly shows the expansion of the field with height, as expected.

Detecting linear polarization signals in $\lambda$8542 is more challenging. However, the total linear polarization ($TLP = \sum_{\lambda}\sqrt{Q_{\lambda}^2+U_{\lambda}^2}/I_{\lambda}$) clearly differentiates between plage patches ("footpoints") with weak linear polarization, and the fibril canopy with relatively stronger signals. In some instances, we observe polarization features aligned with the direction of the fibrils (Fig.\,\ref{fig:TLPTCP1}\textcolor{xlinkcolor}{a}), with an example indicated by the white arrow in Fig.\,\ref{fig:TLPTCP1}\textcolor{xlinkcolor}{c}. Previous observations did not reveal these features as clearly \citep[e.g.,][]{2020A&A...642A.210M,2020A&A...644A..43P}, likely due to higher noise.

\begin{figure}[ht]
    \centering
    \includegraphics[width=\linewidth]{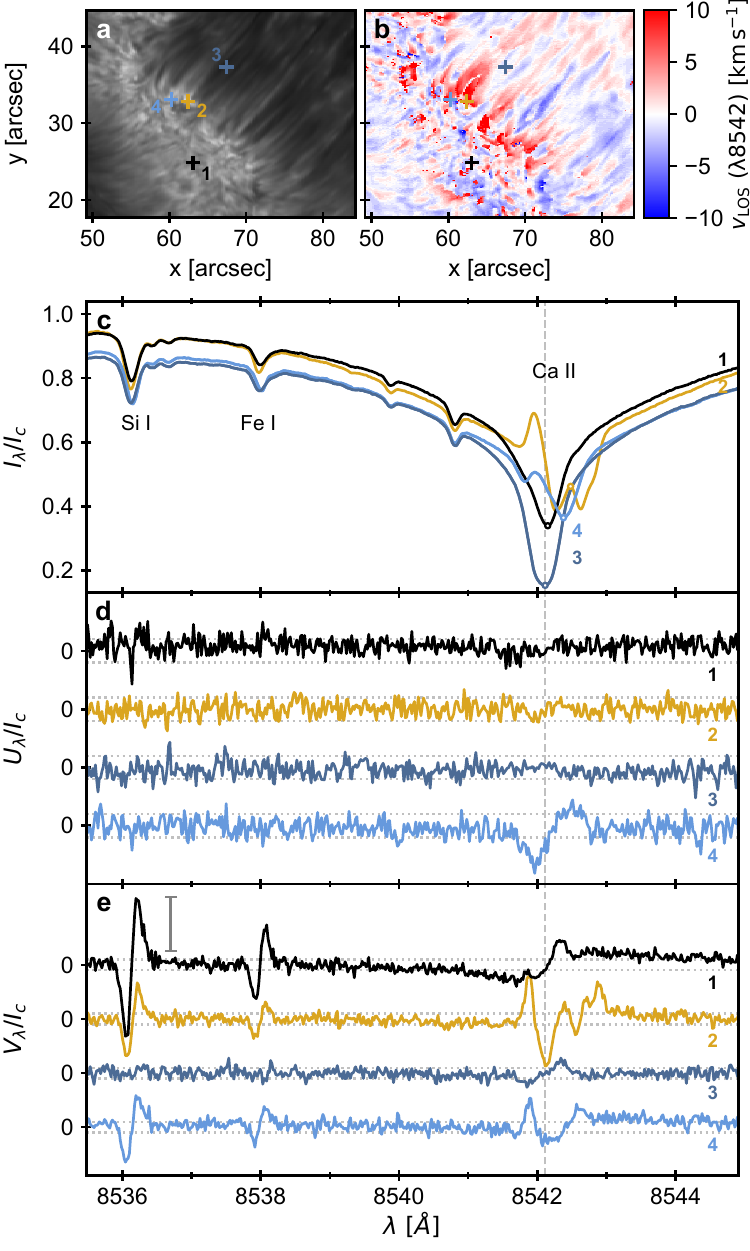}
    \caption{Flows and polarization in and around plage. Panels a-b: zoomed-in views in ($\lambda$8542) intensity/dopplergram of the region delimited by the white box in Fig. \ref{fig:TLPTCP1}; the color bar range is capped. Panels c-e: Stokes I, U, and V signals at the locations marked by the crosses in the top panels (no spatial binning). The vertical dashed line shows the nominal line center. The open circles in panel c show the positions used to determine $V_{\rm LOS}$. Stokes U and V have been vertically shifted for qualitative comparison; the dotted lines show the $\pm2\times10^{-3}$ levels for each spectrum; the vertical bar in panel e indicates an amplitude of $V/I_{\rm c}=2\times10^{-2}$.}    
    \label{fig:TLPTCP2}
\end{figure}

In the core of plages, we find the well-known $\lambda$8542 profiles with higher core intensities (e.g., black line in Fig. \ref{fig:TLPTCP2}\textcolor{xlinkcolor}{c}), but they are not as flat as previously reported \citep[e.g.,][]{2013ApJ...764L..11D,2020A&A...644A..43P}, possibly due to differences in the target or the higher spectral resolution of ViSP. 
The brightest pixels are found at the footpoints of fibrils (e.g., yellow cross in Fig.\,\ref{fig:TLPTCP2}\textcolor{xlinkcolor}{a}). The line profiles at those locations are asymmetric, often showing large Doppler shifts and multiple peaks, indicating strong velocity gradients and supersonic ($>$\,7\,$\rm km\,s^{-1}$) downflow velocities up to $\sim$15\,$\rm km\,s^{-1}$ (Fig.\,\ref{fig:TLPTCP2}\textcolor{xlinkcolor}{b,c}). The Doppler shifts in the plage interiors are well-balanced without a clear preference for upflows or downflows. We find no evidence for supersonic flows in the photosphere (based on ME inversions), but that may require higher spatial resolution and height-dependent inversions \citep{2019A&A...630A..86B}.

While Stokes Q and U signals are generally weak in the FOV, they are clearly detectable at the periphery of the plage regions (Fig.\,\ref{fig:TLPTCP2}\textcolor{xlinkcolor}{d}). Thanks to 
%ViSP's fine spectral sampling, 
an effective spectral resolution of $\sim$\,$1.2\times10^5$,
we observe complex, multi-lobe profiles in the corresponding Stokes V signals (Fig.\,\ref{fig:TLPTCP2}\textcolor{xlinkcolor}{e}). Those profiles generally reproduce the derivative of the intensity, as expected under the WFA. We find significant Stokes V signals throughout the entire FOV, sometimes stronger in the chromosphere than in the photospheric \ion{Si}{I} and \ion{Fe}{I} lines seen in the far wings (Fig.\,\ref{fig:TLPTCP2}\textcolor{xlinkcolor}{e}); this occurs at locations where the fibrils begin to extend outward from the photospheric footpoints. 

\begin{figure*}[ht]
    \centering
    \includegraphics[width=\linewidth]{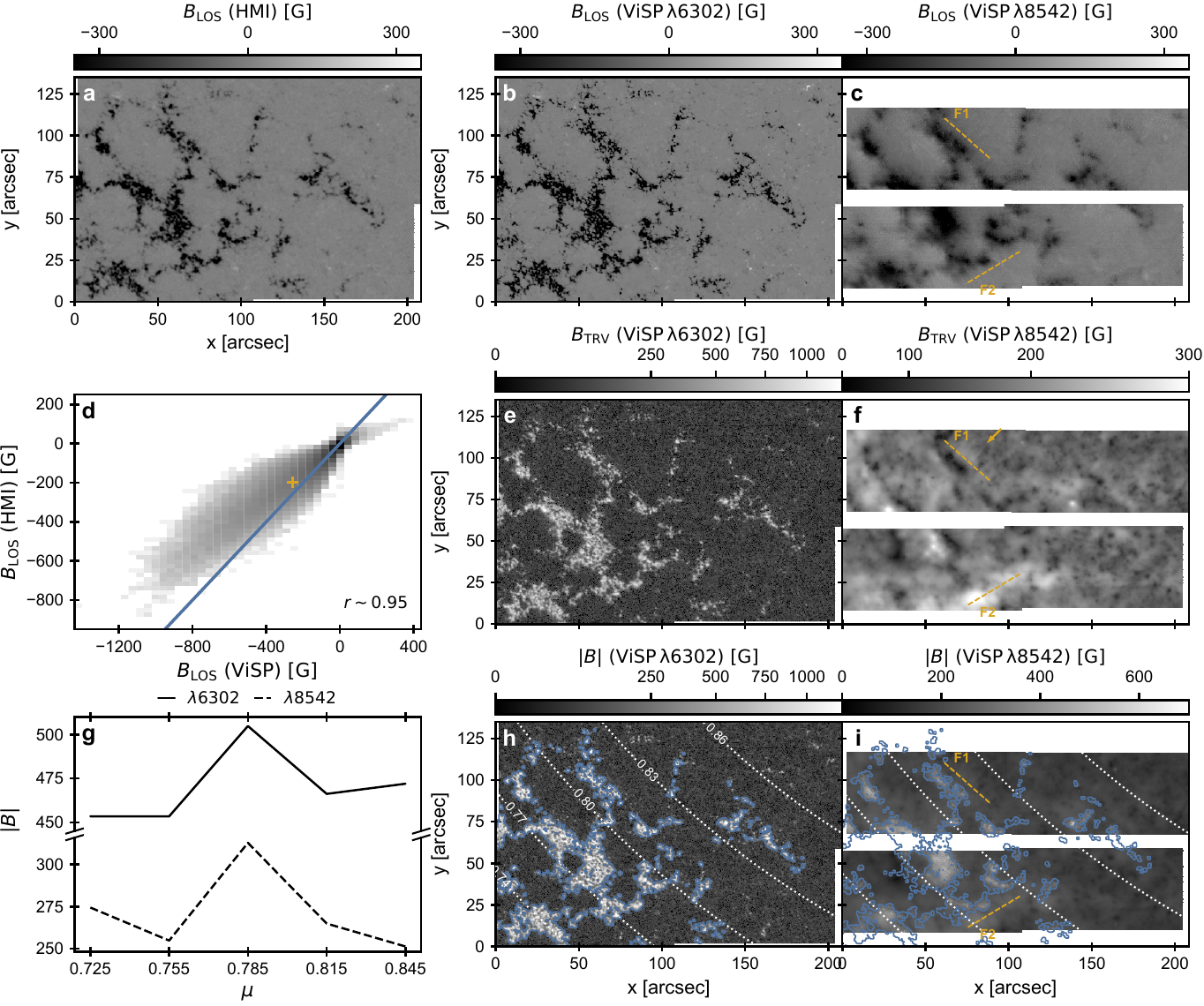}
    \caption{Magnetic field in the photosphere and chromosphere. Panel a: synchronic HMI LOS magnetogram, matched to  spatiotemporal location of ViSP slit. Panel b-c: LOS component of the field from ViSP $\lambda6302$ and $\lambda8542$. Panel d: two-dimensional histogram of LOS magnetic field values derived from ViSP vs HMI; the solid lines shows the HMI=ViSP locus; the cross shows the mean values within the plage mask (panel h); $r$ is the Pearson's correlation coefficient. Panels e-f: transverse components of the field from ViSP; the arrow highlights the fibril-like structure of the field. Panel g: averaged field strength in plage as a function of $\mu$ for the two ViSP diagnostics. Panels h-i: total field strength from ViSP; the dotted lines show isocontours of $\mu$ values; the blue contours show the plage mask (see text); the lines F1 and F2 are perpendicular to the fibrils (c.f. Fig.\,\ref{fig:TLPTCP1}). The color bars are capped for display purposes. Panels e, f, and h have been gamma adjusted.}
    \label{fig:CLV}
\end{figure*}

\subsection{Stratification of the magnetic field}

Figure \ref{fig:CLV} shows the results of the inversions. To directly comparing the ViSP raster scans with HMI's imaging LOS magnetograms, we cross-correlated the continuum intensity at each ViSP scan position with the closest-in-time HMI image. Using each spatial alignment, we constructed a synchronic HMI magnetogram (Fig. \ref{fig:CLV}\textcolor{xlinkcolor}{a}) for the full ViSP four-pointing mosaic. The qualitative agreement between the two is high (correlation coefficient $r=0.95$), but the ViSP magnetogram shows more fine structure and stronger fields relative to HMI, as expected (Fig. \ref{fig:CLV}\textcolor{xlinkcolor}{b}, actual range: [-1430, 330] G); this is particularly significant for the minority polarity fields that tend to cluster on small spatial scales unresolved by HMI (Fig. \ref{fig:CLV}\textcolor{xlinkcolor}{d}).
We note that the effective spatial resolution of the ViSP data, due to the slit width, integration time, and binning along the slit, is 0$^{\prime\prime}$.25-0$^{\prime\prime}$.45.
We estimated the uncertainty in $B_{\rm LOS}$ (photosphere) to be quite low on the order of $\sim$2\,G. Still, we only see few opposite polarity field patches around plage at this spatial resolution, except at a few locations (e.g., top left in Fig. \ref{fig:CLV}\textcolor{xlinkcolor}{b}). Over the whole FOV, the detected total positive polarity flux is only about 4\% of the negative polarity flux. 

The field weakens and becomes more space-filling in the chromosphere (Fig. \ref{fig:CLV}\textcolor{xlinkcolor}{c}), with an average magnitude of $B_{\rm LOS}$$\sim$\,$-$210\,G ($|B|\,$$\sim\,$280\,G) in plage, while still showing a fair amount of spatial structure. When zooming in on Fig. \ref{fig:CLV}\textcolor{xlinkcolor}{c}, the imprint of some fibrils becomes clearly visible; the amplitude of the field variation across the fibrils is relatively small ($<$50\,G) but significant (Appendix \ref{section:supplements}). The transverse field map also shows fibril-like structures surrounding the plage patches (Fig. \ref{fig:CLV}\textcolor{xlinkcolor}{f}). For example, along the line F1 (Fig. \ref{fig:CLV}\textcolor{xlinkcolor}{i}) we find a mean value of $B_{\rm TRV}$\,$\sim$\,180\,G, which is higher than in the neighboring plage ($B_{\rm TRV}$\,$\sim$\,120\,G), resulting in a total field strength of $|B|\,$$\sim\,$200\,G, while in the case of line F2, we find $|B|\,$$\sim\,$270\,G, but a spatially smoother field. We caution that residual crosstalk and other polarization artifacts in $\lambda$8542, especially in Stokes Q and U, might lead to an overestimation of $B_{\rm TRV}$; however, the inferred spatial structure is credible. Supplementary discussion and figures are presented in the Appendix \ref{section:supplements}. 

The center-to-limb (CLV) variation of the field strength can be used to investigate the field expansion with height \citep[e.g.,][]{1997ApJ...474..810M}; the overall FOV, while not optimal, is sufficiently large for such an analysis. 
To this end, we segmented the plage regions using a (photospheric) field strength threshold of 150\,G, which typically defines the plage perimeter very well, regardless of its evolutionary state \citep{2000ssma.book.....S}. We computed the mean field strength in plage elements for different $\mu$ bins with widths of 0.03. 
We excluded from this calculation pores (based on a continuum intensity threshold) and some small ($\lesssim$\,1$^{\prime\prime}$) isolated patches outside the plage (e.g., top and lower right in Fig. \ref{fig:CLV}\textcolor{xlinkcolor}{h}). We find a slight downtrend with decreasing $\mu$, at least in the photosphere (Fig. \ref{fig:CLV}\textcolor{xlinkcolor}{g}).
We find a stronger CLV effect when applying a more restrictive field threshold, though this reduces the statistical significance. The plage region at $\mu\sim0.78$, being relatively strong compared to the other patches, results in a high value in the plot (Fig. \ref{fig:CLV}\textcolor{xlinkcolor}{g}). For the $\lambda$8542, the same trend is not visible, likely because of a reduced sample size (due to data gaps) and inherently higher uncertainty of the field strength ($\sim$10\%, Appendix\,\ref{section:supplements}). 

%\vfill\null % to force columnbreak

\subsection{Field strength and heating rates}

The magnetic field is believed to play a key role in chromospheric plage heating, either through magnetohydrodynamic waves, reconnection, electric current dissipation, or ion-neutral effects \citep[e.g.,][]{2019ARA&A..57..189C}. \citet{2021ApJ...921...39A} reported a weak correlation between the field strength from \ion{He}{I} 10830 and the integrated intensities in the \ion{Mg}{II} h and k lines. In contrast, \citet{2021SciA....7.8406I}, using the same line (\ion{Mg}{II} k) to derive both intensities and longitudinal field strengths, found a strong correlation between the two quantities. However, this correlation weakened in deeper layers.

Wavelength-integrated \ion{Ca}{II} K intensities have been shown to correlate with the radiative losses \citep[][]{2021A&A...647A.188D}. Here, we used $\lambda8542$ intensities since they generally correlate with \ion{Ca}{II} K \citep[see also ][]{2018A&A...612A..28L}, as a proxy for the heating rates, leveraging the large FOV compared to previous observations. We find a clear correlation between the longitudinal field strength and the integrated intensities (within $\pm$200\,$\rm m\mathring{A}$) in the FOV ($r=0.76$); however, the correlation degrades when we apply the plage mask ($r=0.64$, see Fig. \ref{fig:CLV}\textcolor{xlinkcolor}{i}) or when using the total field strength ($r<0.6$).  

% ========================================================% 
\section{Discussion and Conclusions}

This study focused on characterizing the dynamic environment and magnetic field stratification of an extended plage region using high signal-to-noise spectropolarimetric data obtained by DKIST/ViSP. 
 
The analysis shows strong circular polarization signals in both the photosphere and chromosphere of plages. We detected weak but discernible linear polarization signals, differentiating between plage patches and the magnetic canopy, showing fibril-like polarization features coaligned with the fibrils. 

Plages still appear as predominantly unipolar regions in the photosphere at the low noise level of the ViSP magnetograms ($\sim$2\,G), with a low prevalence of opposite polarity flux ($\sim$4\% in the whole FOV). We find little evidence of opposite polarities at the plage peripheries \citep[e.g., unlike in the AR plage in][]{2017ApJS..229....4C} or plage interiors as implicated by the apparent small-scale loop-like structures seen in AIA EUV images \citep[e.g.,][]{2016ApJ...820L..13W,2019ApJ...885...34W}. 
%Other high-resolution ($\sim$\,0$^{\prime\prime}.1-0^{\prime\prime}$.3) observations have not detected a significant contribution from opposite polarity flux within plages \citep[e.g.,][]{2015A&A...576A..27B,2020A&A...642A.210M}, but it is unclear what lies at higher resolution. 

The average magnitude of the LOS chromospheric magnetic field in this plage is approximately $-210$\,G, significantly smaller than previously reported values \citep[$B_{\rm LOS}\sim-420$\,G, ][]{2020A&A...642A.210M}, perhaps explained by differences in the  targets and observing angles. However, our average value is consistent with the range of middle-chromospheric values derived by the CLASP2 observations for similar $\mu$ values \citep{2021SciA....7.8406I}. %The total field strength averages to $\sim\,$280\,G in plage.

The LOS field distribution across the fibrils exhibits small ($<$50 G) but clear variations, showing fibril-like structures surrounding the plage patches; while this is expected, previous observations could not show it as clearly. In fibrils, we find total field strengths between approximately $200-300$\,G. The lower limit is compatible with the field strengths implied by the transverse wave properties observed in chromospheric fibrils rooted in the network \citep{2017A&A...607A..46M}, though such wave properties could differ in plage. The higher limit is consistent with the field strength in fibrils reported by \citet{2020A&A...644A..43P} based on observations at a similar $\mu$. However, their maps showed little spatial structure. % {\bf While an improved ViSP polarization calibration is needed to reduce systematic uncertainties, particularly in $B_{\rm TRV}$, the inferred spatial structure is credible.}
%LOS field values of a few hundred gauss have also been reported in off-limb spicules \citep{2020A&A...642A..61K}.

We investigated the relationship between the longitudinal field strength and the intensities in the $\lambda$8542 line, a proxy for heating rates. A clear correlation was observed when considering the whole FOV, but the correlation weakens in plages, indicating more complex interactions between the magnetic field and heating, or no connection between them. It has been suggested that resistive heating at the edges of flux tubes in the lower chromosphere can give rise to enhanced $\lambda$8542 emission offset from the magnetic footpoints \citep{2013ApJ...764L..11D,2022A&A...664A...8M}, in which case we do not expect a correlation with B but with the curl of B. The radiative cooling rate in the \ion{Ca}{II} 8542 line is as strong as in the \ion{Ca}{II} K line in the QS \citep{1981ApJS...45..635V}, but it remains to be quantified how well the former serves as a proxy for the heating rates in plages. Other effects, such as magneto-acoustic shocks, may play a role in the chromosphere of plages, but their study requires higher temporal cadence observations.

We find evidence of supersonic (up to $\sim$\,15\,$\rm km\,s^{-1}$) downflows in the chromosphere at the plage periphery. \citet{2022A&A...661A.122S} found supersonic downflows ($\sim$\,20$\pm7\rm\,km\,s^{-1}$) at plages boundaries using \ion{He}{I} 10830\,\AA~GREGOR/GRIS observations. DKIST observations reveal that these downflows can be traced to relatively deeper layers
%as we expect the \ion{He}{I} 10830\,\AA~line to be formed higher in the atmosphere 
\citep[e.g., Fig. 7 in ][]{2017SSRv..210..109D}. These flows might be attributed to cold material draining from the fibrils along highly inclined fields, interleaved with the fibril field, causing local heating and enhanced emission where shocks form \citep[see ][]{2007A&A...462.1147L, 2011ApJ...742..119R}. A superposition of flows with different velocity components possibly account for the multi-lobed intensity profiles. In that case, the WFA may not be adequate. Non-LTE inversions will provide further insight into the thermodynamics of these regions.

Our findings highlight the importance of considering the magnetic structure and the dynamic environment when investigating the heating mechanisms within plages. Further research with higher spatial/temporal resolution and a larger sample size at more heliocentric angles could provide insights into the height expansion of the magnetic field, the relationship between the magnetic field strength/topology and heating, and the specific mechanisms driving the dynamics and heating in plage regions. 
Extended chromospheric magnetograms will be of interest for field extrapolations into the heliosphere for comparison with in situ PSP measurements.

\begin{acknowledgments}
We thank Christian Beck, Momchil Molnar, and Ivan Milić for the useful discussions on the ViSP data. The research reported herein is based in part on data collected with the Daniel K. Inouye Solar Telescope (DKIST), a facility of the National Solar Observatory (NSO). NSO is managed by the Association of Universities for Research in Astronomy, Inc., and is funded by the National Science Foundation. DKIST is located on land of spiritual and cultural significance to Native Hawaiian people. The use of this important site to further scientific knowledge is done so with appreciation and respect. Any opinions, findings and conclusions or recommendations expressed in this publication are those of the author(s) and do not necessarily reflect the views of the National Science Foundation or the Association of Universities for Research in Astronomy, Inc.  
\end{acknowledgments}


\vspace{5mm}
\facilities{DKIST(ViSP \& VBI), SDO(AIA \& HMI)}


\software{\texttt{Astropy} \citep{2013A&A...558A..33A,2018AJ....156..123A,2022ApJ...935..167A}, \texttt{Sunpy} \citep{sunpy_community2020},
          \texttt{pyMilne} \citep{2019A&A...631A.153D},
          \texttt{spat\_WFA} \citep{2020A&A...642A.210M},
          \texttt{adhoc\_xtalk} \citep{2022ApJ...930..132J}
          }



%\newpage
\appendix

\section{Data reduction and post-processing}
\label{section:post_processing}

\subsection{Data reduction}
\label{section:reduc}
The data pipeline producing Level 1 ViSP data includes dark/bias signal subtraction, gain correction, geometric calibration, background light calibration, and polarization calibration. We refer to the calibration documentation\footnote{\url{https://docs.dkist.nso.edu/projects/visp/en/v2.0.1/index.html}} for further details.
%\label{section:xtalk}

\subsection{Crosstalk correction}
\label{section:crosstalk}
We removed residual crosstalk from Stokes I to Q, U, and V in the Level 1 ViSP data using the approach of \citet{2022ApJ...930..132J}. This method essentially infers the Mueller matrix, $\mathbb{M}$, that transforms the Stokes vector that enters the optical path of the telescope, $\mathbf{S}$, into the measured Stokes vector, $\mathbf{S}^{\prime}$, after passing through all the optical elements ($\mathbf{S}^{\prime}=\mathbb{M}\mathbf{S}$); $\mathbf{S}$ can be obtained by applying the inverse of the Mueller matrix to the measured Stokes vectors. While this is done to a great extent during the Level 1 data reduction using a theoretical Mueller matrix, some residual crosstalk remains. The Mueller matrix can be decomposed into a general diattenuator and a general elliptical retarder ($\mathbb{M}=M_{P}M_{R}$). We note that the method as presented \citet{2022ApJ...930..132J} assumes that the polarization properties are spatially uniform and constant in wavelength, which we found not to be true for this ViSP data set, as the crosstalk changes along the slit and in wavelength. 
Therefore, we have adapted the code\footnote{\url{https://github.com/sajaeggli/adhoc_xtalk}} to derive the diattenuation matrix along the slit by averaging the spectra along the scan direction. 
The diattenuation matrix, $M_{P}$, is defined as
\begin{equation}
\begin{split}
M_{P} &=
\begin{pmatrix}
1 & d_{H} & d_{45} & d_{R} \\
d_{H} & \sqrt{1-D^{2}} & 0 & 0 \\
d_{45} & 0 & \sqrt{1-D^{2}} & 0 \\
d_{R} & 0 & 0 & \sqrt{1-D^{2}} 
\end{pmatrix} + \frac{1-\sqrt{1-D^{2}}}{D^{2}}
\begin{pmatrix}
0 & 0 & 0 & 0 \\
0 & d^2_{H} & d_{45}d_{H} & d_{R}d_{H} \\
0 & d_{H}d_{45} & d^2_{45} & d_{R}d_{45} \\
0 & d_{H}d_{R} & d_{45}d_{R} & d^2_{R} 
\end{pmatrix},\\
d_{H} &=D\cos\alpha\sin\beta, \\
d_{45} &=D\sin\alpha\sin\beta, \\
d_{R} &=D\cos\beta,
\end{split}
\end{equation}
\noindent where $D$ is the magnitude of the diattenuation vector, and $\alpha$ and $\beta$ are the equivalent angles for the vector; those three parameters are determined such that the correlations of Stokes Q, U, and V to I are minimized \citep[Eq. 16 in][]{2022ApJ...930..132J}. 
While this approach is more computationally intensive than standard methods \citep[e.g.,][]{1992ApJ...398..359S}, we find it to effectively remove the remaining I\,$\rightarrow$\,Q, U, V crosstalk. 

For the 6302\,\AA~arm, we find approximately constant polarization across the wavelength range spanned by the \ion{Fe}{I} 6301 and 6302\,\AA~lines, so the wavelength dependency can be ignored. For the 8542\,\AA~arm, the polarization changes near the core of the $\lambda$8542. The polarization dependency with wavelength may come from the effects of parasitic spectral-dependent scattered light that was known to be present in this early DKIST data. Therefore, we split the crosstalk correction for two different wavelength bins to remedy this.  

We have also tried to remove the V\,$\leftrightarrow$\,Q, U crosstalk using the optimizer for the retardance matrix, $M_{R}$, provided by \citet{2022ApJ...930..132J}, but the code did not converge, possibly because the polarization signals are not strong enough. Nonetheless, that contribution is expected to be significantly smaller than I\,$\rightarrow$\,Q, U, V.

\begin{figure}[ht]
    \centering
    \includegraphics[width=\linewidth]{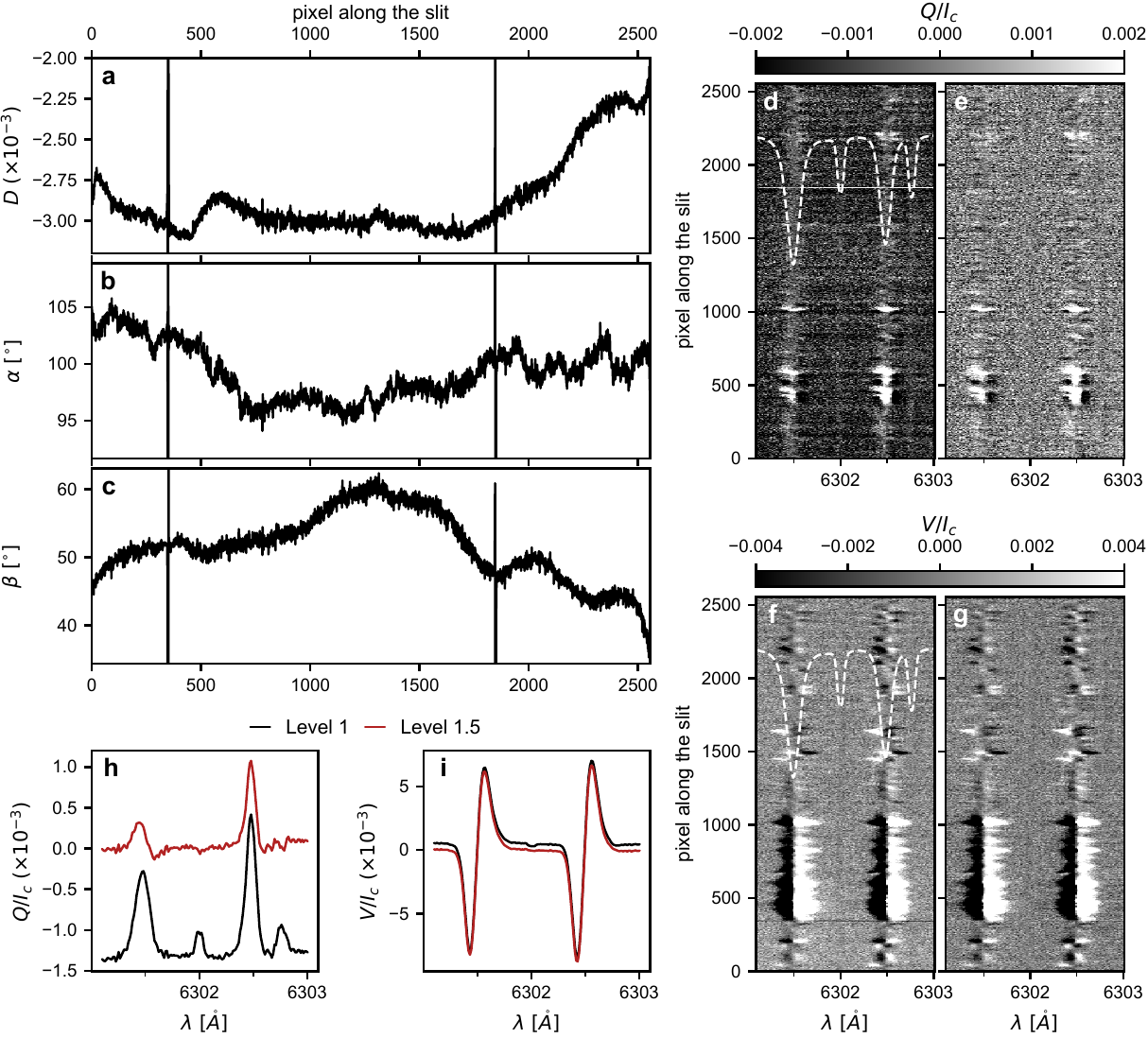}
    \caption{Crosstalk correction for the ViSP spectra. Panels a-c: diattenuation matrix parameters along the ViSP slit for one of the four mosaic tiles. Panels d-g: Level 1 (left) and Level 1.5 (right) Stokes Q and V for a given slit position; the dashed lines show the averaged Stokes I around the \ion{Fe}{I} lines (arbitrary units) for reference; the color bars are capped for display purposes. Panels h-i: averaged Stokes Q and V along the slit shown in panels d-g, with (red) and without (black) crosstalk correction.}
    \label{fig:xtalk}
\end{figure}

Figure \ref{fig:xtalk} illustrates the results of the I\,$\rightarrow$\,Q, U, V crosstalk correction for one of the four ViSP mosaic tiles. We find a significant variation of the diattenuation matrix parameters along the slit (Fig. \ref{fig:xtalk}\textcolor{xlinkcolor}{a-c}). The spikes around x$=$300 and 1800 correspond to the fiducial marks on the slit. Accounting for the measured spatial variation of the Mueller matrix across the FOV removes most of the residual crosstalk in Q, U, and V (examples in Fig. \ref{fig:xtalk}\textcolor{xlinkcolor}{d-g}). Spatially averaged Stokes Q, U, and V in the Level 1 data show a polarization imprint of the telluric lines and a continuum polarization offset, both of which are removed after post-processing (Fig. \ref{fig:xtalk}\textcolor{xlinkcolor}{h-i}). The magnitude of the correction averages to approximately a factor of two for Stokes V and a factor of eight for Stokes Q and U in the continuum at 6302\,\AA.

\subsection{Polarization and intensity residuals}
\label{section:polresiduals}

The ViSP data exhibit additional artifacts that pose challenges in their removal. These artifacts include polarization streaks along the slit of the same or higher magnitude than the real signals, which result from residual intensity crosstalk uncorrected by dual-beam combination, as well as horizontal fringe-like patterns with varying (spatial) frequency along the slit. The former is more prevalent in arm 8542\,\AA~than for arm 6302\,\AA. The streaks have a quasi-periodicity of 8 spatial pixels; therefore we partly removed them using two-dimensional Fourier filtering by masking their corresponding frequencies and computing the inverse transforms. We ensured no ringing was introduced in the data. While the horizontal fringes are not clearly observable at individual wavelengths, they become more evident when integrating in wavelength to produce TLP and TCP maps (Section \ref{section:results}). To mitigate their impact on these maps for both ViSP arms, we isolated the fringe pattern by masking the magnetic concentrations, averaging the TLP and TCP along the scan direction, and subtracting the pattern from the maps. A similar approach was employed to remove fringes from the derived $B_{\rm TRV}$ maps from $\lambda8542$; $B_{\rm LOS}$ is not significantly affected by the residuals. 

There is also a mean intensity mismatch in the red wing of the $\lambda$8542 compared to the solar atlas profile on the order of a few percent. We did not attempt any ad hoc correction to this. 

The WFA method demonstrates robustness in handling polarization and intensity residuals by utilizing only the line core intensities and effectively smoothing out isolated bad pixels through spatial regularization. We provide uncertainty estimates in the Appendix \ref{section:supplements}. Nevertheless, future work should aim at improving the calibration of arm 8542\,\AA~prior to more detailed analysis, such as comparisons with synthetic spectra from semi-empirical models and numerical simulations.

\subsection{Wavelength calibration}
\label{section:vlos}

The wavelength calibration was performed by cross-correlation with the solar atlas \citep{1984SoPh...90..205N}. Also through comparison with the atlas, we find the spectral full width at half maximum of these observations to be approximately 70\,m\AA, consistent with the relatively large width of the spectrograph slit.
An additional correction was applied to remove a repeatable trend in spectral drift as a function of slit position. This was done by averaging the wavelength position of the telluric lines for each pixel in the FOV and interpolating the spectra to the same wavelength grid for both ViSP arms. The magnitude of the drift increases along the scan direction by up to $\sim$0.9$\,\rm km\,s^{-1}$, which is a significant fraction of typical photospheric velocities, thus leading to systematic trends in Dopplergrams. In the chromosphere, the drift pattern is less evident because of the higher mean velocities there (Fig. \ref{fig:TLPTCP2}). 

\subsection{Mosaicking and arm coalignment}

The coordinate information provided in the ViSP FITS file headers is not designed to be sufficiently accurate for correctly mosaicking the four different pointings. We determined the spatial shifts by performing cross-correlation analysis on the overlapping regions of consecutive pointings. Additionally, we achieved co-alignment between the two ViSP arms by cross-correlating the continuum intensities at the same pixel scale, in agreement with the calculations of atmospheric differential refraction based on \citet{2006SoPh..239..503R}. We note that the spectra acquired in both arms are taken simultaneously.

\section{Supplementary figures}
\label{section:supplements}

\begin{figure}[ht]
    \centering
    \includegraphics[width=\linewidth]{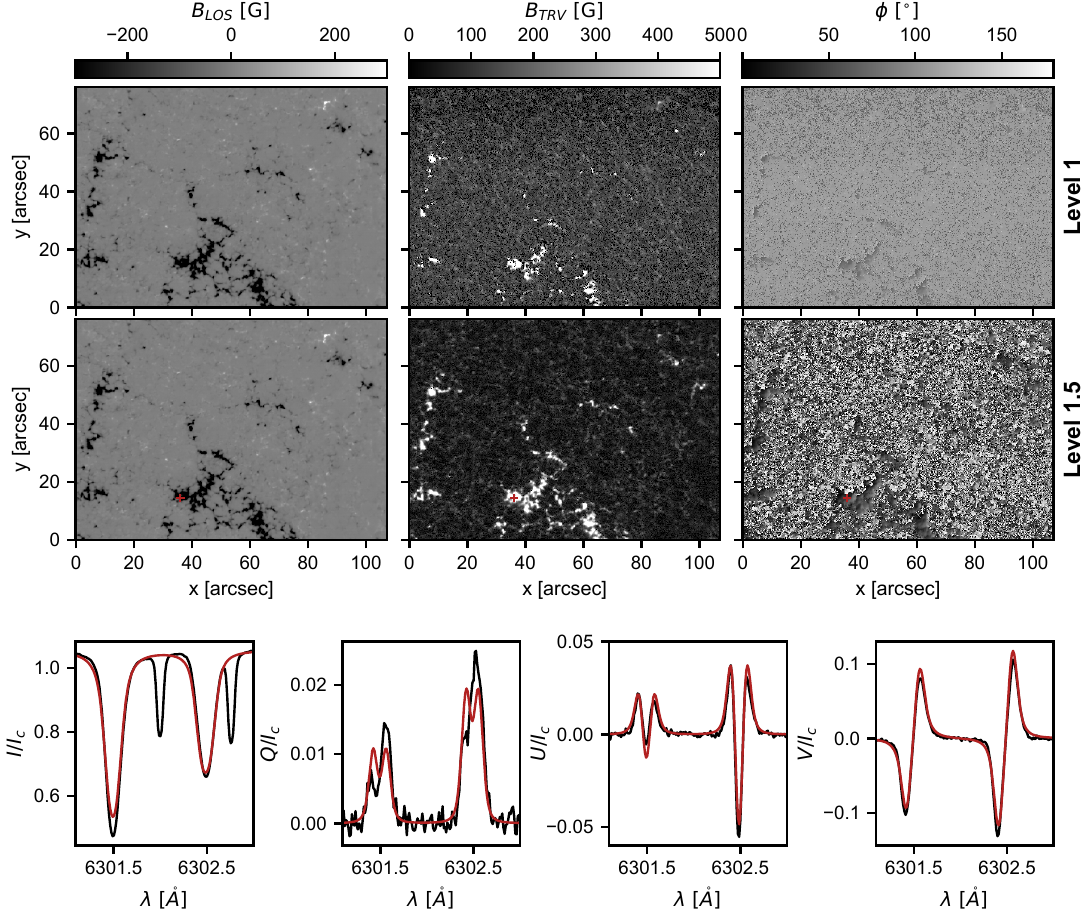}
    \caption{Effect of crosstalk removal in the determination of the photospheric magnetic field. Top panels: Line-of-sight component of the magnetic field (left), transverse component (middle), and azimuth angle (right) for one of the four ViSP mosaic tiles without any post-processing (top row) and with crosstalk removal (bottom row). The color bars are capped for display purposes. Bottom panels: an example observed (Level 1.5) full Stokes spectrum (black) and best fit (red) at the location marked by the red cross in the upper panels. } 
    \label{fig:fieldsxtalk}
\end{figure}

Figure \ref{fig:fieldsxtalk} compares the results of the ME inversions ($\lambda$6302) using the Level 1 and Level 1.5 data. Naturally, the correction is strongest in the magnetic azimuth angle and the transverse component of the field, for which, in particular, an average QS field of $B_{\rm TRV}$\,$\sim$\,130\,G reduces to $B_{\rm TRV}$\,$\sim$\,60\,G after post-processing. In plage regions, the correction in the field strength is typically on the order of a few percent, but it can be as much as $\sim$60\% in some small ($<$\,0$^{\prime\prime}$.5) patches at the plage edges. We find similar corrections for the WFA applied to $\lambda8542$ (not shown). 
The quality of the ME inversion fits is generally satisfactory, as exemplified in the lower panels. Intensity residuals could potentially be due to velocity or temperature gradients and non-LTE effects beyond the scope of the ME approximation \citep[e.g.,][]{2020A&A...633A.157S}. However, gain-correction errors (Section\,\ref{section:reduc}) could not be ruled out. Stokes V is typically better reproduced than Stokes Q and U. 

\begin{figure*}
    \centering
    \includegraphics[width=0.5\linewidth]{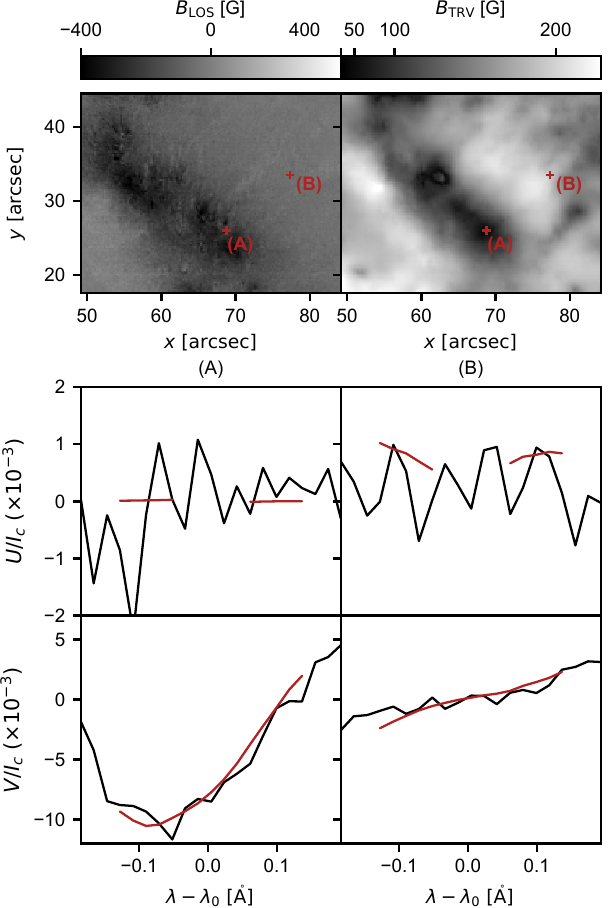}
    \caption{ Magnetic field components in the chromosphere. Top panels: line of sight component (left) and transverse component (right) of the magnetic field in the region shown in the top panels in Fig.\,\ref{fig:TLPTCP2}; $B_{\rm TRV}$ has been gamma-adjusted to enhanced the visibility of the fibrils (c.f. Fig.\,\ref{fig:TLPTCP1} and Fig.\,\ref{fig:TLPTCP2}). Lower panels: example observed Stokes U and V (black lines) and WFA fit (red lines; within $\pm$120\,$\rm m\mathring{A}$ from line nominal line center) at the two locations marked in the upper panels.} \label{fig:WFA_fit}
\end{figure*}

Figure \ref{fig:WFA_fit} presents a closer look at the chromospheric magnetic field encompassing the plage region within the upper portion of the FOV displayed in Fig.\,\ref{fig:TLPTCP2}. Comparison between observed $\lambda$8542 Stokes V profiles and the synthesis from the WFA ($V_{\lambda}\propto B_{\rm LOS}\,dI_{\lambda}/d\lambda$) shows a good correspondence between the two both in plage and fibrils. It is difficult to judge the quality of the fit based on individual Stokes Q or U profiles given the noise; however, it is noteworthy that the WFA code employs spatial regularization to infer a transverse field value that best matches a neighborhood of pixels, helping to mitigate noise and artifacts in any given pixel. Wavelength offsets ($\lambda-\lambda_{0}$) smaller than the Doppler width of the line were excluded from the Stokes Q and U fit \citep[refer to][]{2020A&A...642A.210M}, hence the gaps in the red curves. 
Fitting residuals in the Stokes parameters for both lines ($\lambda$6302 and $\lambda$8542) may be attributed to data acquisition anomalies, including vibrations in the optical elements, and other inaccuracies in polarization calibration (Section\,\ref{section:polresiduals}) that could not be rectified during post-processing (Section\,\ref{section:crosstalk}).

\begin{figure*}
    \centering
    \includegraphics[width=0.75\linewidth]{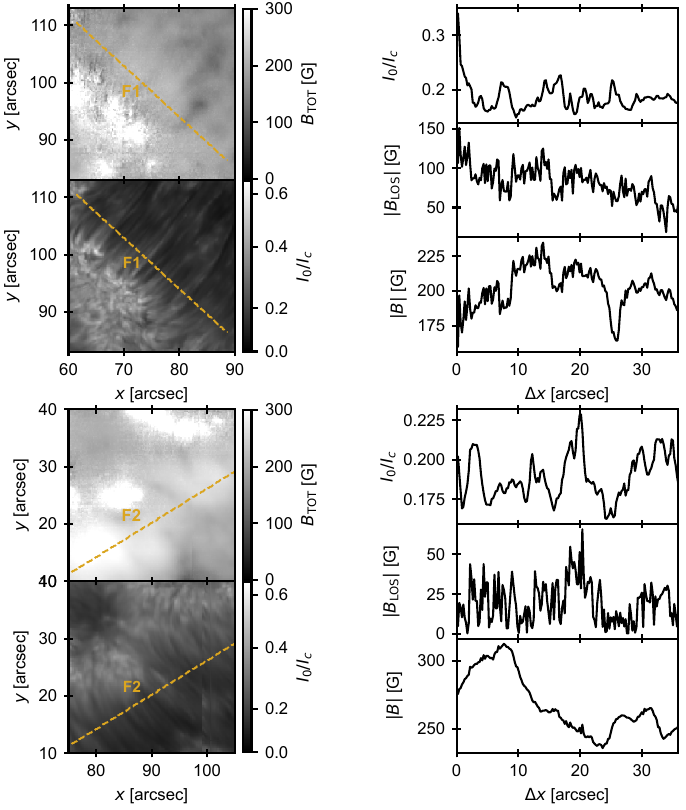}
    \caption{ Magnetic field and $\lambda$8542 intensity across the fibrils around plage regions. Left panels: enlarged views of the regions around the F1 and F2 lines shown in Fig.\,\ref{fig:CLV} in $\lambda$8542 core intensity and total field strength (chromosphere); the colorbars are capped for display purposes. Right panels: intensity, absolute value of the LOS component, and total field strength along the dashed lines shown on the left panels.}  \label{fig:fibrils}
\end{figure*}

Figure \ref{fig:fibrils} shows the variation of the intensity and magnetic field strength along the F1 and F2 slices, as indicated in Fig.\,\ref{fig:CLV}. Despite the fibril-like appearance of the magnetic field around plage, the correlation between the $\lambda$8542 intensities and the field strength along those slices appears nonlinear, transitioning from positive to negative at distinct locations. However, we reiterate the smoothing effect of the regularized WFA in the field components. 
Based on the standard deviation of the inversion results for different reasonable regularization weights, we have estimated the uncertainty in $B_{\rm LOS}$ to be on the order of 1\%~in plages. However, this uncertainty increases away from the magnetic field concentrations; for example, the median uncertainty is 5\%~along the line F2. This means that the observed lateral variations of $\sim$20\,G over spatial scales of $\sim$1$^{\prime\prime}$ are significant (lower right panels in Fig.\,\ref{fig:fibrils}). The uncertainty further rises to a few tens of percent in isolated pixels in the quiet areas, which are not the focus of this paper. The total field strength distribution along F2 is smoother than along F1 likely due to less favorable seeing conditions when the ViSP was scanning that region, as evidenced by the intensity images. The uncertainty in $B_{\rm TRV}$ is around 5\%~in fibrils, increasing to $\sim$10-40\%~in the plage cores where Stokes Q and U are weak. Nonetheless, because the magnetic field strength in plages is generally dominated by the LOS component, the uncertainty in the total field strength remains below $\sim$10\%. While the statistical uncertainties are fairly small, we cannot rule out sources of systematic uncertainties due to the limitations of the current ViSP data calibration (Appendix \ref{section:post_processing}).

\clearpage

%\bibliography{main}{}
\bibliographystyle{aasjournal}



\end{document}